\documentclass[pra,twocolumn,amsmath,amssymb,letterpaper]{revtex4} 
\usepackage{graphicx}
\usepackage{amsmath}
\usepackage{amsfonts}
\usepackage{amssymb}
\usepackage{hyperref}

\begin{document}

\title{Landau's criterion for an anisotropic Bose-Einstein condensate}

\author{Zeng-Qiang Yu}
\email{zqyu.physics@outlook.com}
\affiliation{Institute of Theoretical Physics, Shanxi University, Taiyuan 030006, China}

\begin{abstract}
In this work, we discuss the Landau's criterion for anisotropic superfluidity. To this end, we consider a point-like impurity moving in a uniform Bose-Einstein condensate with either interparticle dipole-dipole interaction or Raman induced spin-orbit coupling. In both cases, we find that the Landau critical velocity $v_{\rm c}$ is generally smaller than the sound velocity in the moving direction. Beyond $v_{\rm c}$, the energy dissipation rate is explicitly calculated via a perturbation approach. In the plane-wave phase of a spin-orbit coupled Bose gas, the dissipationless motion is suppressed by the Raman coupling even in the direction orthogonal to the recoil momentum. Our predictions can be tested in the experiments with ultracold atoms.
\end{abstract}

\maketitle

One of the manifestations of superfluidity is the ability to support dissipationless motion.  An impurity immersed in a superfluid can move without friction when its velocity is below a critical value. This phenomenon is explained by the famous Landau's criterion~\cite{LevSandroBook,LeggettBook}, according to which,
the critical velocity is given by
\begin{align}
  v_{\rm c} = \min \left( \frac{E_{\bf q}}{q} \right), \label{Eq_vc}
\end{align}
with $E_{\bf q}$ being the elementary excitation spectrum of the superfluid. The Landau critical velocity had been observed in liquid He-II and Bose-Einstein condensate (BEC) of a dilute gas, where the onset of dissipation is due to the creation of roton~\cite{He_exp} and phonon~\cite{vc_exp1}, respectively.

Recently, two new types of BEC were realized in the ultracold atomic gases: One was achieved in magnetic atoms of $^{52}$Cr~\cite{Cr_BEC}, $^{164}$Dy~\cite{Dy_BEC} and $^{168}$Er~\cite{Er_BEC} with interparticle dipole-dipole interaction (DDI); the other was achieved in alkali atoms of $^{87}$Rb~\cite{NIST2011} and $^{23}$Na~\cite{Ketterle2016} with synthetic spin-orbit~(SO) coupling. Remarkably, the excitation spectra of these superfluids are anisotropic and can be tuned by an external magnetic field or laser field. While anisotropic dynamics and collective excitations have been experimentally observed in these systems~\cite{dipole_exp1,dipole_exp2,dipole_exp3,dipole_exp4,dipole_exp5, soc_exp1,soc_exp5,soc_exp6}, the critical velocity has not been measured so far.

The application of the Landau's criterion to an anisotropic superfluid is not a trivial problem. In this work, we show that the anisotropic critical velocity, along a certain direction $\bf \hat v$, is given by
\begin{align}
  v_{\rm c}  = \min \left( \frac{E_{\bf q}}{\bf q \cdot \hat v} \right), \label{Eq_vc2}
\end{align}
where the minimum is calculated over all the values of momentum $\bf q$ under the constraint ${\bf q\cdot \hat v}>0$. The condition $v<v_{\rm c}$, with $v_{\rm c}$ given by~(\ref{Eq_vc2}), can be recognized as a generalized Landau's criterion for anisotropic superfluidity, and its  consequence is discussed  later in both dipolar and SO coupled Bose gases.

Critical velocity~(\ref{Eq_vc2}) can be derived in the standard way similar to the isotropic case~\cite{LeggettBook}. Consider a heavy point-like impurity moving in a uniform superfluid at zero temperature. The impurity can lose its kinetic energy by creating one or more excitations in the superfluid, and the onset of dissipation only involves the processes of creating a single excitation~\cite{note9}. Such processes conserve energy and momentum, thus, satisfy the relation
\begin{align}
  {\bf q \cdot v} = E_{\bf q}, \label{conservation_law}
\end{align}
where $\bf v$ is the velocity of the impurity, $\bf q$ is the momentum of the excitation, and we have assumed that the mass of impurity is very large compared to that of atoms in the superfluid. If  condition~(\ref{conservation_law}) cannot be satisfied by any $\bf q$, no excitation will be created. As a result, when $v<v_{\rm c}$, the motion (in the $\bf \hat v$ direction ) is dissipationless.

The difference between~(\ref{Eq_vc}) and (\ref{Eq_vc2}) is not insignificant. In an isotropic system, the minimum value of $E_{\bf q}/({\bf q\cdot \hat v})$ always appears at a momentum in the $\bf \hat v$ direction, consequently, $v_{\rm c}$ reduces to its well-known form~(\ref{Eq_vc}). In an anisotropic system, however, the minimization in~(\ref{Eq_vc2}) is complicated, and in general, $v_{\rm c}$ is not determined by the dispersion relation in the moving direction (see examples below). Physically, (\ref{Eq_vc}) and (\ref{Eq_vc2}) correspond to distinct scenarios of creating excitations at the threshold velocity. In an isotropic system, the excitation created at $v_{\rm c}$ propagates in the same direction as $\bf \hat v$. In contrast, in an anisotropic system, the excitation could propagate in a different direction other than $\bf \hat v$. Geometrically, for a given $\bf\hat v$, one can determine $v_{\rm c}$ as follows: Plot the set of curves of $E_{\bf q}$ versus ${\bf q} \cdot {\bf \hat v}$ along different directions of momentum; then, from the origin, draw tangent lines to each of the curves and find the line with the smallest positive slope; this slope is just equal to $v_{\rm c}$.

For the system with Galilean invariance, the condition $v<v_{\rm c}$ also ensures the stability of a supercurrent state, as first proposed by Landau~\cite{LevSandroBook,LeggettBook}. Suppose a fluid flows at a constant velocity in the $-\bf \hat v$ direction. The appearance of an excitation at momentum $\bf q$ costs an energy $E_{\bf q}+{\bf q}\cdot {\bf v}_{\rm flow}$. Therefore, the flow suffers an energetic instability once $v_{\rm flow}$ exceeds $v_{\rm c}$. It should be noted that if the Galilean invariance is broken, the maximum velocity of the stable superflow would be different from the critical velocity~(\ref{Eq_vc2}).

Beyond $v_{\rm c}$, the motion of the impurity becomes dissipative. To calculate the energy dissipation rate, we assume the interaction between the impurity and the condensate is weak and treat it as a perturbation (set $\hbar=1$),
\begin{align}
  \mathcal{H}_{\rm pert} = \sum_{j=1}^N U ({\bf r}_j - {\bf v} t) = \int \frac{{\rm d}^3 q}{(2\pi)^3}\, \rho_{\bf q}^\dagger U_{\bf q} e^{-i{\bf q}\cdot {\bf v}t}
\end{align}
where $U({\bf r})$ is the interaction potential between the impurity and the atoms of the superfluid, $\rho_{\bf q}=\sum_{j=1}^N e^{-i{\bf q}\cdot {\bf r}_j}$ is the density fluctuation operator,  $N$ is the atoms number, and $U_{\bf q}=\int {\rm d}^3r\, U({\bf r})e^{-i{\bf q}\cdot{\bf r}}$. For convenience, we take $U({\bf r})$ as a contact potential and set $U_{\bf q}=U_0$. The dissipation rate of the impurity $P$ equals to the energy transferred to the condensate in the per unit time $d\mathcal{E_{\rm SF}}/dt$. In the linear response regime, we have~\cite{LevSandroBook}
\begin{align}
  P =  \frac{ U_0^2n}{4\pi^2N} \int {\rm d}^3q\int {\rm d}\omega \, \omega S({\bf q},\omega) \delta (\omega-{\bf q}\cdot {\bf v}) \label{dissipation}
\end{align}
where $n$ is the average density of atoms, and $S({\bf q},\omega)= \left| \langle \Phi_{\bf q} | \rho_{\bf q}^\dag | \Phi_0 \rangle \right|^2 \delta(\omega-E_{\bf q})$ is the dynamic structure factor of the superfluid, with $|\Phi_0\rangle$ and $|\Phi_{\bf q}\rangle$ being the ground state and the excited state respectively. Since $S({\bf q},\omega)$ is non-zero only at the resonant frequencies $\omega=E_{\bf q}$, the energy dissipation vanishes for $v<v_{\rm c}$, which is consistent with the Landau's criterion. In experiments, the dissipation can be observed through the heating of the superfluid.

The dissipation rate~(\ref{dissipation}) can be also written as $ P = -{\bf F} \cdot {\bf v}$,
where ${\bf F}=U_0^2n/(4\pi^2N) \int {\rm d}^3 q \,  S({\bf q},{\bf q}\cdot {\bf v}) {\bf q}  $ is the drag force acting on the impurity~\cite{Pitaevskii2004}. The drag force arises for $v>v_{\rm c}$. In an isotropic superfluid, from a symmetry argument, one can show that $\bf F$ is always in the direction of $-\bf \hat v$. This conclusion no longer holds in an anisotropic system~\cite{note_drag}. Nevertheless, only the projection $\bf F\cdot \hat v$ contributes to the energy dissipation.

Now we apply the above general results to a dipolar BEC~\cite{note_dipole}. The interparticle interactions in a dipolar gas consist of two parts: the short range van der Waals interaction and the long-range DDI. Assuming  all the dipoles aligned along the $\bf \hat z$ axis, the effective interaction potential reads~\cite{dipole_review}
\begin{align}
 \mathcal{V}_{\rm int}({\bf r}) = g \left[ \delta({\bf r}) +  3\epsilon_{\rm dd}  (1-3\cos^2\theta)/(4\pi r^3) \right],
\end{align}
where $m$ is the atomic mass, $g$ is the coupling constant of the $s$-wave contact interaction, $\theta$ is the angle between the dipole axis and the interatomic distance $\bf r$, and $\epsilon_{\rm dd}$ is a dimensionless parameter quantifying the relative strength of the DDI. In experiments, $\epsilon_{\rm dd}$ can be tuned via the technique of Feshbach resonance.

The Bogoliubov excitation spectrum of a polarized dipolar BEC is given by~\cite{dipole_review}
\begin{align}
  E_{\bf q} &= q \sqrt{q^2/(4m^2) + c^2(\theta_{\bf q})}, \label{Eq_dipole}
\end{align}
where $c(\theta_{\bf q})$ is the anisotropic sound velocity depending on the polar angle of the momentum $\bf q$,
\begin{align}
   c(\theta_{\bf q}) &= \sqrt{ c^2_\parallel \cos^2\theta_{\bf q} + c^2_\perp \sin^2\theta_{\bf q} }. \label{sound_velocity}
\end{align}
The sound velocity reaches its minimum $c_\perp=c_0\sqrt{1-\epsilon_{\rm dd}}$ at $\theta_{\bf q}=\pi/2$ and reaches its maximum $c_\parallel=c_0\sqrt{1+2\epsilon_{\rm dd}}$ at $\theta_{\bf q}=0$ or $\pi$, where  $c_0=\sqrt{gn/m}$ is the sound velocity with a pure $s$-wave interaction. Below, we restrict our discussion in the regime $0<\epsilon_{\rm dd}<1$ and $g>0$ to make sure the excitation spectrum is stable.

\begin{figure}
\includegraphics{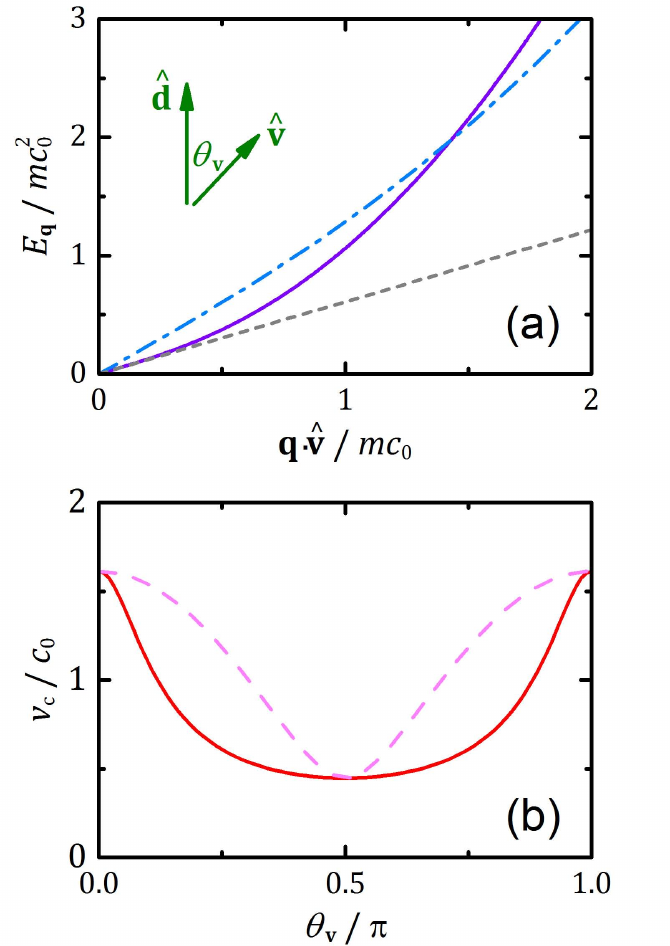}
\caption{(a) Excitation spectrum $E_{\bf q}$ of a dipolar BEC as a function of $\bf q \cdot \hat v$,  where $\bf \hat v$ is in the direction with $\theta_{\bf v}=\pi/4$ (see the inset). The solid line is for the momentum $\bf q$ in the direction with $\theta_{\bf q}$ determined by Eq.~(\ref{theta_dipole}); the dashed-dotted line is for  $\theta_{\bf q}=\theta_{\bf v}$; the dotted line is tangent to the solid line in the $q\rightarrow 0$ limit, and its slope is equal to $v_{\rm c}$.  (b) Anisotropic critical velocity as a function of $\theta_{\bf v}$ (solid line). The dashed line shows the sound velocity in the moving direction. For both plots, $\epsilon_{\rm dd}=0.8$.} \label{Fig_vc_dipole}
\end{figure}

Since the system possesses a rotation symmetry about the dipole axis, the anisotropic critical velocity only depends on the polar angle of the moving direction $\theta_{\bf v}$. From the dispersion relation~(\ref{Eq_dipole}), it is readily shown that the onset of dissipation is due to the creation of a phonon, and the critical velocity  (2) can be written as
\begin{align}
  v_{\rm c}(\theta_{\bf v}) = \min \left[ \frac{c(\theta_{\bf q})}{\cos(\theta_{\bf q}-\theta_{\bf v})} \right], \label{vc_phonon}
\end{align}
where the minimum is calculated over all the values of $\theta_{\bf q}$ under the constraint $\theta_{\bf v}-\pi/2<\theta_{\bf q}<\theta_{\bf v}+\pi/2$. After a straightforward algebra, we find that the phonon created at $v_{\rm c}$ carries a momentum with $\theta_{\bf q}$ satisfying
\begin{align}
  c_\perp^2 \tan \theta_{\bf q} =  c_\parallel^2 \tan \theta_{\bf v} . \label{theta_dipole}
\end{align}
Clearly, except for the special cases of $\theta_{\bf v}=0, \pi/2$ and $\pi$, the propagating direction of the phonon is different from the moving direction of the impurity, in accordance with the scenario we discussed previously.
Substituting (\ref{theta_dipole}) into (\ref{vc_phonon}), we obtain the critical velocity
\begin{align}
  v_{\rm c}(\theta_{\bf v}) = c_\parallel c_\perp \left( \sin^2\theta_{\bf v}c_\parallel^2 + \cos^2\theta_{\bf v}c_\perp^2 \right)^{-1/2}. \label{vc_dipole}
\end{align}
Unlike the condensate with a pure $s$-wave interaction,  the critical velocity of the dipolar BEC is generally smaller than the sound velocity in the moving direction~\cite{note2}
\begin{align}
  v_{\rm c} (\theta_{\bf v}) \leqslant c(\theta_{\bf v}),
\end{align}
where the equality holds only for the motion in the direction either parallel or perpendicular to the dipole axis. The difference between the values of $v_{\rm c}$ and $c$ is illustrated in Fig.~\ref{Fig_vc_dipole}. For a given $\epsilon_{\rm dd}$, the ratio $v_{\rm c}/c$ takes the minimum value at $\theta_{\bf v}=\pi/4$.

\begin{figure}
\includegraphics{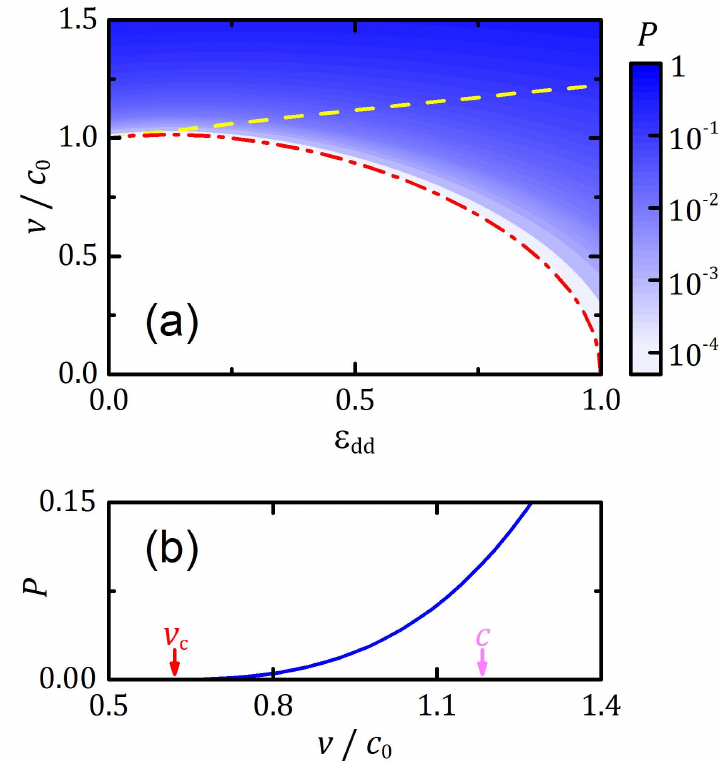}
\caption{(a) Dissipation rate $P$ of a moving impurity in a polarized dipolar BEC as a function of $v$ and $\epsilon_{\rm dd}$. The dash-dotted line shows the critical velocity; and the dashed line shows the sound velocity in the moving direction. (b) Velocity dependence of $P$ for $\epsilon_{\rm dd}=0.8$. For both plots, $\theta_{\bf v} = \pi/4$, and $P$ is in units of $U_0^2nm^3c_0^3$.} \label{Fig_dissipation_dipole}
\end{figure}

Beyond $v_{\rm c}$, the dissipation rate (5) can be written as
\begin{align}
  P = \frac{U_{0}^2 n }{8\pi^2} \int {\rm d}^3q\, \frac{q^2}{m} \delta( E_{\bf q}-{\bf q}\cdot {\bf v}),  \label{dissipation_dipole}
\end{align}
where we have used the $f$-sum rule $\int {\rm d}\omega \, \omega S({\bf q},\omega)=Nq^2/(2m)$ to simplify the expression. For $\epsilon_{\rm dd}=0$, Eq.~(\ref{dissipation_dipole}) recovers the previous result with a pure $s$-wave interaction, $ P = U_0^2 n m^3 (v^2-c_0^2)^2 \Theta(v-c_0)/(\pi v)$~\cite{Pitaevskii2004,Kovrizhin2001}. For $\epsilon_{\rm dd}>0$, the dissipation rate is numerically calculated as a function of velocity $v$, and the results for $\theta_{\bf v}=\pi/4$ are displayed in Fig.~\ref{Fig_dissipation_dipole}. Obviously, the energy dissipation arises even in the subsonic regime ($v<c$), which is in contrast to isotropic weakly interacting bosons. In the limit $\epsilon_{\rm dd}\rightarrow 1$, $v_{\rm c}$ approaches to zero due to the vanishing $c_\perp$. However, in this case, the dissipation rate increases very slowly at small~$v$,  because the phase space contributing to the integral in (\ref{dissipation_dipole}) is quite limited.

Next, we turn to another anisotropic superfluid -- a SO coupled BEC. The synthetic SO coupling had been successfully realized in ultracold atoms using the technique of the two-photon Raman transition~\cite{soc_review}. Here, we consider the spin-half bosons subjected to the equal contributions of  Rashba and Dresselhaus SO coupling. The single-particle Hamiltonian is given by~\cite{NIST2011}
\begin{align}
  \mathcal{H}_{\rm SO} = \frac{({\bf p}-{\bf k}_{\rm r}\sigma_z)^2}{2m} + \frac{\Omega}{2} \sigma_x + \frac{\delta}{2}\sigma_z,
\end{align}
where ${\bf k}_{\rm r}$ is the recoil momentum of the Raman lasers, $\Omega$ is the Raman coupling strength, $\delta$ is the detuning of the Raman transition, and $\sigma_x$ and $\sigma_z$ are the Pauli matrices. Below, we assume ${\bf k}_{\rm r}$ to be oriented along the $\bf \hat z$ axis and set the detuning $\delta$ equal to zero. The term $\sim {\bf p}\cdot {\bf k}_{\rm r}\sigma_z$ represents the SO coupling, which breaks the symmetry of Galilean invariance~\cite{note7}.  The interactions between atoms can be written as $\mathcal{H}_{\rm int}=g/2 \int {\rm d}^3 r\, n^2({\bf r})$, where we have assumed the equal coupling constants in the different spin-channels with $g>0$. Such a choice is relevant to the current experiments with $^{87}$Rb atoms~\cite{NIST2011,soc_exp1,soc_exp5,soc_exp6}, in which the spin-dependent interaction is very weak.

The ground state phase diagram of a SO coupled Bose gas has been studied previously~\cite{NIST2011,Ho,Trento1}. When the interparticle interaction is spin-independent, the ground state is the plane-wave (PW) phase for $\Omega<\Omega_{\rm c}$ and the zero-momentum (ZM) phase for $\Omega>\Omega_{\rm c}$, where $\Omega_{\rm c}=4E_{\rm r}$, and $E_{\rm r}=k_{\rm r}^2/(2m)$ is the recoil momentum. In the PW phase, the ground state spontaneously breaks the Z$_2$ symmetry, and the condensate occupies a non-zero momentum with a finite spin magnetization. Without loss of generality, we assume the magnetization to be positive. In the ZM phase, bosons condense at ${\bf p}=0$ with a balanced spin population. The transition between these two phases is of second-order nature with a divergent  magnetic susceptibility at the critical point $\Omega_{\rm c}$~\cite{Trento3}.

The excitation spectrum of a SO coupled BEC consists of two branches~\cite{Trento2,Zhengwei}, and the critical velocity is determined by the lower branch. In the long wave-length limit, the lower branch excitation is a phonon mode with anisotropic sound velocity. Remarkably, the orientation dependence of the sound velocity is similar to the dipolar BEC, and it can be written in the form of~(\ref{sound_velocity})~\cite{note3}. In the PW phase, $c_\parallel=c_0\sqrt{1-\Omega^2/\Omega_{\rm c}^2}$; in the ZM phase, $c_\parallel=c_0\sqrt{1-\Omega_{\rm c}/\Omega}$; and in both phases, $c_\perp=c_0$ is independent of $\Omega$. Following the procedure as shown before, we find the critical velocity obtained in~(\ref{vc_dipole}) is also valid for the SO coupled BEC as long as the onset of dissipation is due to the emission of a phonon. This is indeed the case of the ZM phase, and it is also true in the PW phase when $\Omega$ is close to $\Omega_{\rm c}$. At the critical point $\Omega_{\rm c}$, $c_\parallel$ vanishes, and $v_{\rm c}$ equals zero in all the directions except $\theta_{\bf v}= \pi/2$~\cite{note4}, which is similar to the situation of the dipolar BEC when $\epsilon_{\rm dd}\rightarrow 1$~\cite{note5}.

\begin{figure}
\includegraphics{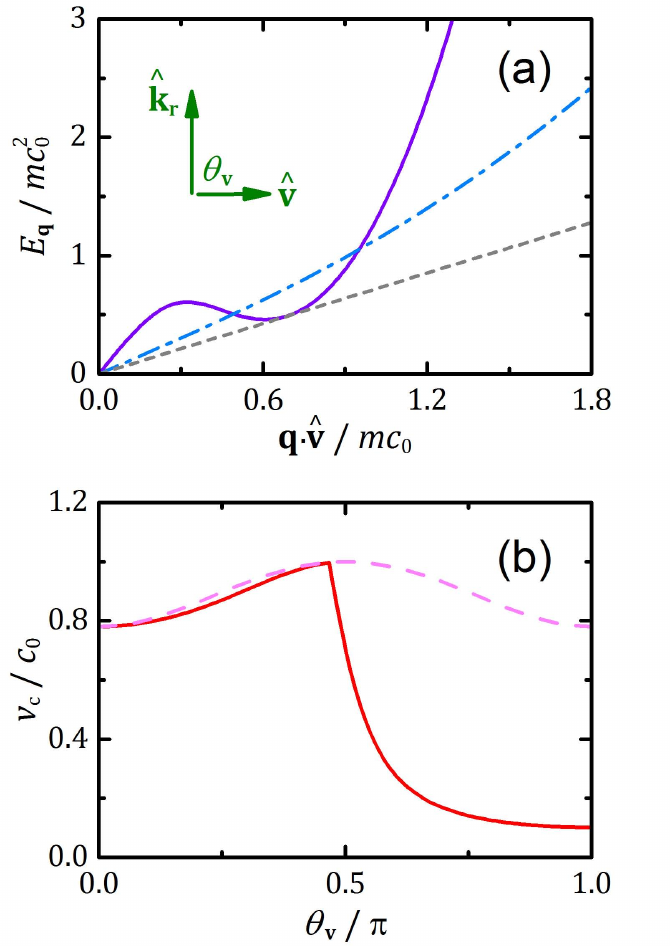}
\caption{(a) Lower branch excitation spectrum $E_{\bf q}$ in the PW phase of a SO coupled Bose gas as a function of $\bf q \cdot \hat v$, where $\bf \hat v$ is in the direction with $\theta_{\bf v}=\pi/2$ (see the inset). The solid line is for the momentum $\bf q$ in the direction that $E_{\bf q}/({\bf q\cdot \hat v})$ reaches the global minimum with $\theta_{\bf q}\simeq 0.91\pi$; the dashed-dotted line is for  $\theta_{\bf q}=\theta_{\bf v}$; the dotted line is tangent to the solid line at the roton minimum, and its slope is equal to $v_{\rm c}$.  (b) Anisotropic critical velocity as a function of $\theta_{\bf v}$ (solid line). The dashed-line shows the sound velocity in the moving direction. For both plots, $\Omega=2.5E_{\rm r}$, and $gn=1E_{\rm r}$.}  \label{Fig_vc_soc}
\end{figure}

For the smaller values of $\Omega$, roton-like excitations emerge~\cite{Trento2,Zhengwei,Higbie} and give rise to a dramatic influence on the anisotropic superfluidity. As an example, in Fig.~\ref{Fig_vc_soc}(a), we show how the critical velocity in the direction perpendicular to ${\bf k}_{\rm r}$ is affected by the roton structure. The non-monotonic behavior of $E_{\bf q}$ only appears for $\theta_{\bf q}$ close to $\pi$. For the motion in the direction with a small $\theta_{\bf v}$, $v_{\rm c}$ is not affected by the roton structure because the minimization in~(2) is under the constraint ${\bf q}\cdot {\bf \hat v}>0$. In Fig.~\ref{Fig_vc_soc}(b), $v_{\rm c}$ is plotted as a function of $\theta_{\bf v}$ for $\Omega=2.5E_{\rm r}$, and there is a kink at $ \theta_\star \simeq 0.47\pi$. For $\theta_{\bf v}<\theta_\star$, $v_{\rm c}$ is determined by the sound velocity of the phonon [Eq.~(\ref{vc_dipole})]; for $\theta_{\bf v}>\theta_\star$, $v_{\rm c}$ is determined by the dispersion near the roton.

\begin{figure}[t]
\includegraphics{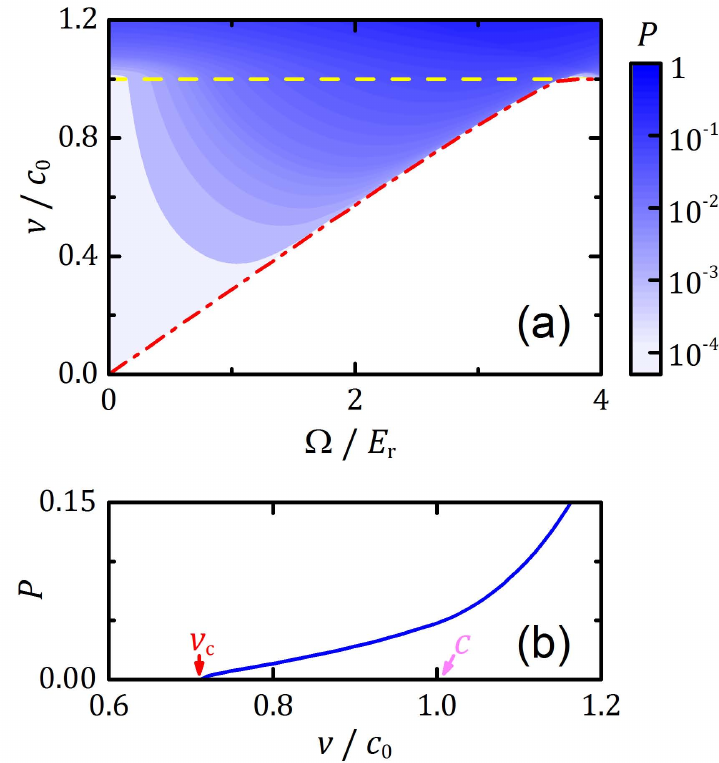}
\caption{(a) Dissipation rate $P$ of a moving impurity in a SO coupled BEC as a function of $v$ and $\Omega$. The dash-dotted line shows the critical velocity; and the dashed line shows the sound velocity in the moving direction. (b) Velocity dependence of $P$ for $\Omega=2.5E_{\rm r}$. For both plots, $\bf \hat v$ is perpendicular to ${\bf k}_{\rm r}$ (i.~e. $\theta_{\bf v} = \pi/2$), $gn=1E_{\rm r}$, and $P$ is in units of $U_0^2nm^3c_0^3$. }  \label{Fig_dissipation_soc}
\end{figure}

Beyond $v_{\rm c}$, the dissipation rate~(\ref{dissipation}) is numerically  calculated within the Bogoliubov approximation~\cite{note6}. Remarkably, when the roton-like excitation emerges, the dissipationless motion is suppressed by the Raman coupling even in the direction orthogonal to ${\bf k}_{\rm r}$, although the excitation spectrum in this direction remains unchanged.  As shown in Fig.~\ref{Fig_dissipation_soc}, for the intermediate values of $\Omega$, visible dissipation arises at velocities between $v_{\rm c}$ and $c_0$. For $\Omega\lesssim 0.5E_{\rm r}$, the dissipation in the subsonic regime becomes negligibly small. This is because, in this case, the roton exhibits a dominant spin-flip nature, and it produces a very weak response to the density fluctuation [see Eq.~(\ref{dissipation})]. In the limit $\Omega\rightarrow 0$, while $v_{\rm c}$ vanishes due to the closing of the roton gap, dissipation only arises for $v>c_0$~\cite{note8}.

Finally, we make some remarks on the normal fluid density, which concerns another aspect of superfluidity. For He-II and ordinary Bose gases, the dissipationless motion below $v_{\rm c}$ is associated with the fact that normal fluid density $\rho_{\rm n}$ vanishes at zero temperature. More precisely, a finite value of $v_{\rm c}$ rules out the existence of low-energy excitations of transverse nature~\cite{LevSandroBook}. This conclusion also holds for the anisotropic systems investigated here. On the other hand, for a SO coupled BEC, the high-energy branch of the excitations has a nontrivial contribution to the transverse current response function, which gives rise to a finite $\rho_{\rm n}$ even at zero temperature~\cite{soc_normal_density1,soc_normal_density2}.

In summary, in this work, we discussed the Landau's criterion for anisotropic superfluidity and investigated its consequence in both dipolar and SO coupled Bose gases. Our predictions can be tested in the experiments with ultracold atoms by moving a microscopic object through a condensate~\cite{vc_exp1}. The general result for critical velocity can be also applied to fermionic superfluids.

The author thanks Hui Zhai, Wei Yi, Bo Liu and Lan Yin for helpful discussions. This work is supported by NSFC under Grant No. 11674202.


\begin{thebibliography}{99}

\bibitem{LevSandroBook}
L. P. Pitaevskii and S. Stringari, {\it Bose-Einstein Condensation} (Oxford University Press, New York, 2003).

\bibitem{LeggettBook}
A. J. Leggett, {\it Quantum Liquids} (Oxford University Press, New York, 2006).

\bibitem{He_exp}
D.~R.~Allum, P.~V.~E. McClintock, A.~Phillips, and R.~M.~Bowley, Philos. Trans. R. Soc. London A {\bf 284}, 179 (1977).

\bibitem{vc_exp1}
A. P. Chikkatur, A. G\"{o}rlitz, D.~M.~Stamper-Kurn, S.~Inouye, S.~Gupta, and W.~Ketterle, Phys. Rev. Lett. {\bf 85}, 483 (2000).

\bibitem{Cr_BEC}  
A. Griesmaier, J.~Werner, S.~Hensler, J.~Stuhler, and T.~Pfau, Phys. Rev. Lett. {\bf 94}, 160401 (2005).

\bibitem{Dy_BEC}  
M.~Lu, N.~Q.~Burdick, S.~H.~Youn, and B.~L.~Lev, Phys. Rev. Lett. {\bf 107}, 190401 (2011).

\bibitem{Er_BEC}  
K. Aikawa, A.~Frisch, M.~Mark, S.~Baier, A.~Rietzler, R.~Grimm, and F.~Ferlaino, Phys. Rev. Lett. {\bf 108}, 210401 (2012).

\bibitem{NIST2011}
Y.-J. Lin, K. Jim\'{e}nez-Garc\'{\i}a, and I. B. Spielman, Nature (London) \textbf{471}, 83 (2011).

\bibitem{Ketterle2016}
J.~Li, W.~Huang, B.~Shteynas, S.~Burchesky, F.~\c{C}.~Top, E.~Su, J.~Lee, A.~O.~Jamison, and W.~Ketterle, Phys. Rev. Lett. {\bf 117}, 185301 (2016).

\bibitem{dipole_exp1}  
T.~Lahaye, T.~Koch1, B.~Fr\"{o}hlich, M.~Fattori, J.~Metz, A.~Griesmaier, S.~Giovanazzi, and T.~Pfau, Nature (London) {\bf 448}, 672 (2007).

\bibitem{dipole_exp2}
T.~Koch, T.~Lahaye, J.~Metz, B.~Fr\"{o}hlich, A.~Griesmaier, and T.~Pfau, Nat. Phys. {\bf 4}, 218 (2008).

\bibitem{dipole_exp3}
T. Lahaye, J. Metz, B. Fr\"{o}hlich, T.~Koch, M.~Meister, A.~Griesmaier, T.~Pfau, H.~Saito, Y.~Kawaguchi, and M.~Ueda, Phys. Rev. Lett. {\bf 101}, 080401 (2008).

\bibitem{dipole_exp4}
G. Bismut, B. Pasquiou, E. Mar\'{e}chal, P.~Pedri, L.~Vernac, O.~Gorceix, and B.~Laburthe-Tolra, Phys. Rev. Lett. {\bf 105}, 040404 (2010).

\bibitem{dipole_exp5}  
G.~Bismut, B.~Laburthe-Tolra, E.~Mar\'{e}chal, P.~Pedri, O.~Gorceix, and L.~Vernac, Phys. Rev. Lett. {\bf 109}, 155302 (2012).

\bibitem{soc_exp1}
J.-Y.~Zhang, S.-C.~Ji, Z.~Chen, L.~Zhang, Z.-D.~Du, B.~Yan, G.-S.~Pan, B.~Zhao, Y.-J.~Deng, H.~Zhai, S.~Chen, and J.-W. Pan, Phys. Rev. Lett. {\bf 109}, 115301 (2012).

\bibitem{soc_exp5}
M. A. Khamehchi, Y. Zhang, C. Hamner, T. Busch, and P. Engels, Phys. Rev. A {\bf 90}, 063624 (2014).

\bibitem{soc_exp6}
S.-C. Ji, L. Zhang, X.-T. Xu, Z. Wu, Y. Deng, S. Chen, and J.-W. Pan, Phys. Rev. Lett. {\bf 114}, 105301 (2015).

\bibitem{note9}
This statement is true in an isotropic superfluid~\cite{LeggettBook}. For the anisotropic superfluids considered here, we take it as an assumption in our calculation.


\bibitem{Pitaevskii2004}
G.~E.~Astrakharchik and L.~P.~Pitaevskii, Phys. Rev. A {\bf 70}, 013608 (2004).

\bibitem{Kovrizhin2001}
D.~L.~Kovrizhin and L.~A.~Maksimov, Phys. Lett. A {\bf 282}, 421 (2001).

\bibitem{note_drag}
The noncollinear drag force has been discussed previously in the BECs with Rashba and Weyl SO coupling~\cite{HePeisong,LiaoRenyuan}.

\bibitem{note_dipole}
Critical velocity of a trapped dipolar condensate has been numerically studied in previous works~\cite{dipole_theory2,dipole_theory3}, the physics discussed there is different from this work.

\bibitem{dipole_review}  
M. A. Baranov, Phys. Rep. {\bf 464}, 71 (2008); T.~Lahaye, C.~Menotti, L.~Santos, M.~Lewenstein, and T.~Pfau, Rep. Prog. Phys. {\bf 72}, 126401 (2009).

\bibitem{note2}
Using Eqs.~(10) and (13), it is readily shown that $c^2(\theta_{\bf v})/v_{\rm c}^2(\theta_{\bf v}) = 1+\frac{1}{4}\sin^22\theta_{\bf v} (c_\parallel^2/c_\perp^2 + c_\perp^2/c_\parallel^2-2) \geqslant 1$.

\bibitem{soc_review}
For recent reviews, see, N. Goldman, G. Juzeli\={u}nas, P. \"{O}hberg, and I. B. Spielman, Rep. Prog. Phys. {\bf 77}, 126401 (2014); H. Zhai, Rep. Prog. Phys. {\bf 78}, 026001 (2015).

\bibitem{note7}
The lack of Galilean invariance has been pointed out in~\cite{Zhengwei,Ozawa}. It should be noted that in the direction perpendicular to ${\bf k}_{\rm r}$ the Galilean invariance is still preserved.

\bibitem{Ho}
T.-L. Ho and S. Zhang, Phys. Rev. Lett. {\bf 107}, 150403 (2011).

\bibitem{Trento1}
Y. Li,  L. P. Pitaevskii, and  S. Stringari, Phys. Rev. Lett. \textbf{108}, 225301 (2012).

\bibitem{Trento3}
Y. Li, G. I. Martone, and S. Stringari, Euro. Phys. Lett. {\bf 99}, 56008 (2012).

\bibitem{Trento2}
G. I. Martone, Y. Li, L. P. Pitaevskii, and S. Stringari, Phys. Rev. A {\bf 86}, 063621 (2012).

\bibitem{Zhengwei}
W. Zheng, Z.-Q. Yu, X. Cui, and H. Zhai, J. Phys. B: At. Mol. Opt. Phys. {\bf 46}, 134007 (2013).

\bibitem{note3}
The anisotropic sound velocity can be understood from the effective mass approximation, $c(\theta_{\bf q})=\sqrt{gn/m^*(\theta_{\bf q})}$, with $1/m^*=\cos^2\theta_{\bf q}/m_\parallel^* + \sin^2\theta_{\bf q}^2/m $. For $\Omega<\Omega_{\rm c}$, $m_\parallel^*= m/\sqrt{1-\Omega_c^2/\Omega^2}$; for $\Omega>\Omega_{\rm c}$, $m_\parallel^*=m/\sqrt{1-\Omega_c/\Omega}$~\cite{Zhengwei}.

\bibitem{note4}
While $v_{\rm c}$ has a discontinuity at $\theta_{\bf v}=\pi/2$, the dissipation rate $P$ is still a smooth function of $\theta_{\bf v}$. As $\theta_{\bf v}\rightarrow \pi/2$, $P$ gradually vanishes in the subsonic regime.

\bibitem{note5}
For the dipolar BEC with $\epsilon_{\rm dd}=1$, $c_\perp=0$, and $v_{\rm c}$ vanishes in all directions except $\theta_{\bf v} = 0$ and $\pi$.

\bibitem{Higbie}
J. Higbie and D.~M.~Stamper-Kurn, Phys. Rev. Lett. {\bf 88}, 090401 (2002).

\bibitem{note6}
For simplicity, we assume that the interaction between the impurity and atoms is spin independent.

\bibitem{note8}
In this limit, we reproduce the result for a single component BEC, $ P = U_0^2 n m^3 {(v^2-c_0^2)^2}\Theta(v-c_0)/(\pi v)$.

\bibitem{soc_normal_density1}
Y.-C. Zhang, Z.-Q. Yu, T. K. Ng, S. Zhang, L. Pitaevskii, and S. Stringari, Phys. Rev. A {\bf 94}, 033635 (2016).

\bibitem{soc_normal_density2}
S. Stringari, arXiv:1609.04694.


\bibitem{HePeisong}
P.-S. He, Y.-H. Zhu, and W.-M. Liu, Phys. Rev. A {\bf 89}, 053615 (2014).

\bibitem{LiaoRenyuan}
R. Liao, O. Fialko, J. Brand, and U. Z\"{u}licke, Phys. Rev. A {\bf 93}, 023625 (2016).

\bibitem{dipole_theory2}  
R. M. Wilson, S. Ronen, and J. L. Bohn, Phys. Rev. Lett. {\bf 104}, 094501 (2010).

\bibitem{dipole_theory3}  
C. Ticknor, R. M. Wilson, and J. L. Bohn, Phys. Rev. Lett. {\bf 106}, 065301 (2011).

\bibitem{Ozawa}
T. Ozawa, L. P. Pitaevskii, and S. Stringari, Phys. Rev. A {\bf 87}, 063610 (2013).


\end{thebibliography}
\end{document}